\documentclass[conference]{IEEEtran}
\IEEEoverridecommandlockouts
\usepackage{cite}
\usepackage{amsmath,amssymb,amsfonts}
\usepackage{algorithmic}
\usepackage{graphicx}
\usepackage{textcomp}
\usepackage{xcolor}
\def\BibTeX{{\rm B\kern-.05em{\sc i\kern-.025em b}\kern-.08em
    T\kern-.1667em\lower.7ex\hbox{E}\kern-.125emX}}
\begin{document}

\title{

%Enhancing weak annotators in active learning for sound event detection \\(or)\\

From Weak to Strong Sound Event Labels using Adaptive Change-Point Detection and Active Learning

%Strong Sound Event Labeling with Adaptive Change-Point Detection and Active Learning
\thanks{This work was supported by The Swedish Foundation for Strategic Research (SSF; FID20-0028) and Sweden's Innovation Agency (2023-01486).
\\\\
%https://github.com/johnmartinsson/active-learning-for-bioacoustics
https://github.com/johnmartinsson/adaptive-change-point-detection
}
}

\author{\IEEEauthorblockN{1\textsuperscript{st} John Martinsson}
\IEEEauthorblockA{\textit{Computer Science} \\
\textit{Research Institutes of Sweden}\\
Gothenburg, Sweden \\
john.martinsson@ri.se}
\and
\IEEEauthorblockN{2\textsuperscript{nd} Olof Mogren}
\IEEEauthorblockA{\textit{Computer Science} \\
\textit{Research Institutes of Sweden}\\
Gothenburg, Sweden \\
olof.mogren@ri.se}
\and
\IEEEauthorblockN{3\textsuperscript{rd} Maria Sandsten}
\IEEEauthorblockA{\textit{Centre for Math. Sciences} \\
\textit{Lund University}\\
Lund, Sweden \\
maria.sandsten@matstat.lu.se}
\and
\IEEEauthorblockN{4\textsuperscript{th} Tuomas Virtanen}
\IEEEauthorblockA{\textit{Signal Proc. Research Centre} \\
\textit{Tampere University}\\
Tampere, Finland \\
tuomas.virtanen@tuni.fi}
}

\maketitle

\begin{abstract}

% Abstract submitted to EUSIPCO 2024
%In this work we propose an audio recording segmentation method based on an adaptive change point detection (A-CPD) for machine guided weak label annotation of audio recording segments. The goal is to maximize the amount of information gained about the temporal activations of the target sounds. For each unlabeled audio recording, we use a prediction model to derive a probability curve used to guide annotation. The prediction model is initially pre-trained on available annotated sound event data with classes that are disjoint from the classes in the unlabeled dataset. The prediction model then gradually adapts to the annotations provided by the annotator in an active learning loop. The queries used to guide the weak label annotator towards strong labels are derived using change point detection on these probabilities. We show that it is possible to derive strong labels of high quality even with a limited annotation budget, and show favorable results for A-CPD when compared to two baseline query strategies.

% Revised abstract
We propose an adaptive change point detection method (A-CPD) for machine guided weak label annotation of audio recording segments. The goal is to maximize the amount of information gained about the temporal activations of the target sounds. For each unlabeled audio recording, we use a prediction model to derive a probability curve used to guide annotation. The prediction model is initially pre-trained on available annotated sound event data with classes that are disjoint from the classes in the unlabeled dataset. The prediction model then gradually adapts to the annotations provided by the annotator in an active learning loop. We derive query segments to guide the weak label annotator towards strong labels, using change point detection on these probabilities. We show that it is possible to derive strong labels of high quality with a limited annotation budget, and show favorable results for A-CPD when compared to two baseline query segment strategies.

%The query segments used to guide the weak label annotator towards strong labels are derived using change point detection on these probabilities.

\end{abstract}


%=======================================================================================================================
% Introduction
%=======================================================================================================================
\section{Introduction}
\label{sec:introduction}
% The introduction should set this paper in the context of sound event annotation, but also the importance of this for bioacoustics and ecouacoustics. 
%My long term goals are broadly to make people at all levels of society more aware of the importance of sound as an information source. And show the potential of machine learning to tap into this information. But, also how important this technology is in bioacoustics, and how pressing the need is for quality annotations in this field. 
%These goals motivate my work, and this should be clear from the introduction. This paper should be one kernel of knowledge in pursuit of this long term goal, and the same context should be part of the coming papers in this active learning for sound series.

Most audio datasets today consists of weakly labeled data with imprecise timing information~\cite{Kong2020}, and there is a need for efficient and reliable annotation processes to acquire labels with precise timing information. We refer to such labels as strong labels. The performance of sound event detection (SED) models improve with strong labels~\cite{Hershey2021}, and strong labels become especially important when we want to count the number of occurrences of an event class. For example in bioacoustics, where counting the number of vocalizations of an animal species can be used to estimate population density and draw ecological insights~\cite{Marques2013}.

%    \item ... is label bias depending on annotators a problem???
%\end{itemize}
%Recent work has suggested that it can be more beneficial to ask annotators for weak labels than strong labels~\cite{Martin-Morato2023a}. 

Crowdsourcing the strong labels is challenging and an attractive solution is to crowdsource weak labels to enable reconstruction of the strong labels~\cite{Martin-Morato2021, Martin-Morato2023a}. Asking the annotator for strong labels requires more work and it can in the worst case lead to the annotator misunderstanding the task~\cite{Martin-Morato2023a}. %\textbf{TODO: add sentence about lack of strong labels, and this other work using weak labels}

\begin{figure}
    \centering
    \includegraphics[width=1.0\columnwidth]{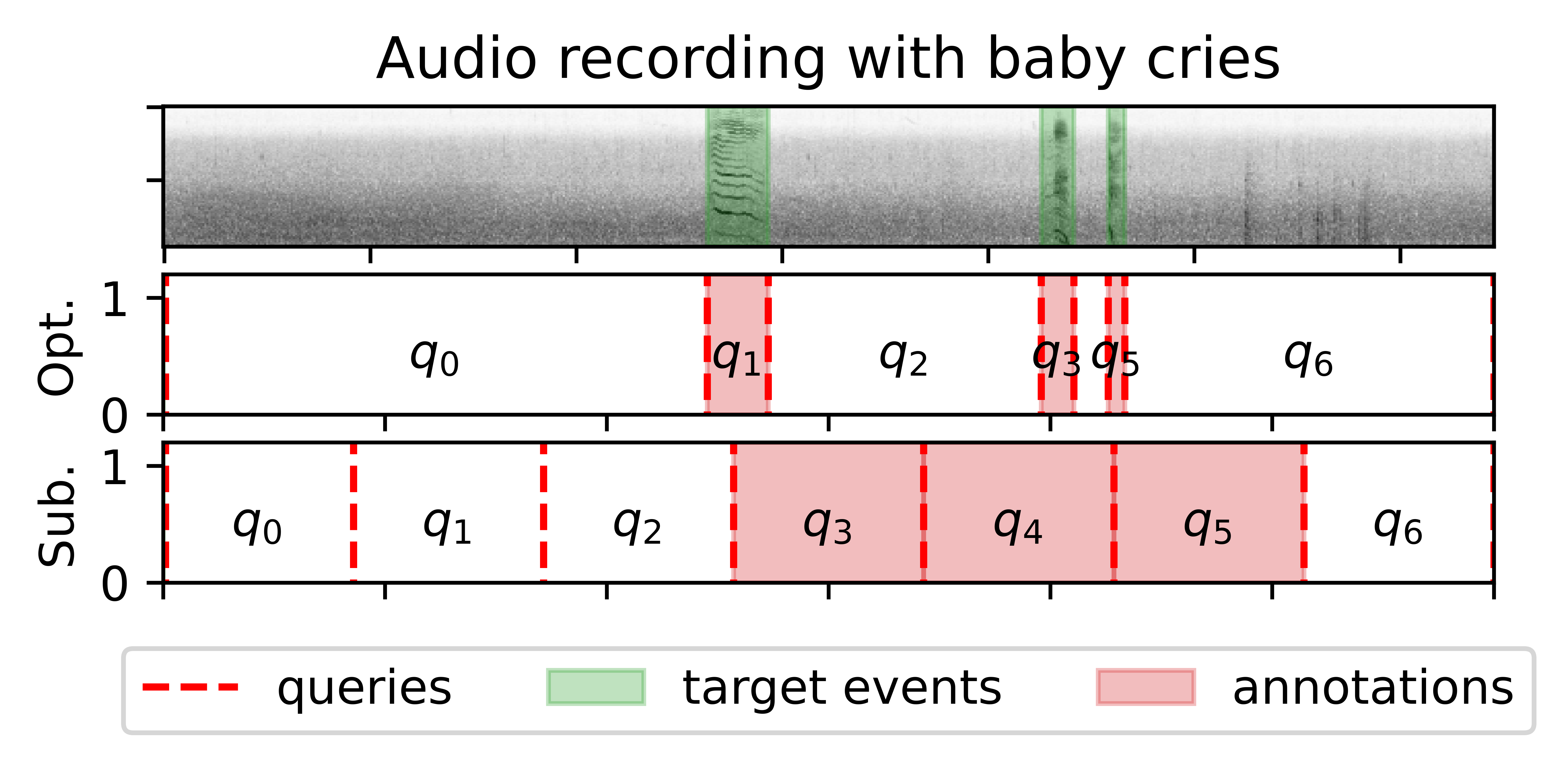}
    \caption{
    Illustration of segmentation of an audio spectrogram with three target events shown in shaded green (top panel) into a set of audio query segments $q_0, \dots, q_6$ using an optimal method w.r.t the derived strong label timings (middle panel) and a sub-optimal method (bottom panel). Resulting annotations, from the weak labels given by the annotator, are shown in shaded red for both methods. Query $q_4$ for the optimal method is omitted for clarity.
    %Queries, $q_1, \dots, q_6$, are illustrated by their start and end time. Asking the optimal queries (Opt.) to the weak label annotator give perfect strong labels, and asking sub-optimal queries (Sub.) results in weaker labels.
    }
    \label{fig:example}
\end{figure}

Disagreement-based active learning is the most used form of active learning for sound event detection~\cite{Shuyang2017, Shuyang2018, Shuyang2020, Wang2022}, focusing on selecting what audio segment to label next. The recordings are either split into equal length audio segments~\cite{Shuyang2017, Shuyang2018, Wang2022} or segments depending on the structure of the sound~\cite{Shuyang2020}. Each segment is then given a weak label by the annotator.

%For a given query budget $B$ the cost of annotating $N$ audio recordings of the same length $T$ become $N(c_1 T + c_2 B)$, where $c_1$ and $c_2$ are the cost constants. Most active learning methods would try to minimize this cost by annotating fewer audio recordings, we instead aim to reduce $B$ while maintaining label quality.

%In this work we use a weak label annotator to derive strong labels as in~\cite{Martin-Morato2023a}, but instead of using fixed length query segments we adapt the query segments to the data to gain more information about the strong label timings. This is done in the setting of active learning and each new annotation improves the query strategy and as a result the overall annotation process. 

%We propose an adaptive change point detection (A-CPD) method which splits a given audio recording into a set of audio segments, or queries. The queries are then labeled by an annotator and the strong labels are derived and evaluated. See Fig.~\ref{fig:example} for an illustration of this concept, where a set of seven queries (the budget) are used either optimally or sub-optimally for a given audio recording. We aim to adapt the set of queries in such a way that the information gained about the strong labels is maximized. 

We use a weak label annotator to derive strong labels as in~\cite{Martin-Morato2023a}, but instead of using fixed length query segments we adapt the query segments to the data, in the setting of active learning. We propose an adaptive change point detection (A-CPD) method which splits a given audio recording into a set of audio segments, or queries. The queries are then labeled by the annotator and the strong labels are derived and evaluated. See Fig.~\ref{fig:example} for an illustration where a set of seven queries are used either optimally or sub-optimally for a given audio recording with three sound events. We assume three sound events to be detected in each audio recording as a simplification during method development. We aim to adapt the set of queries in such a way that the information about the temporal activations of the target sounds is maximized. Note that we aim to actively guide the annotator during the annotation of the audio recordings, rather than actively choose which audio recordings to annotate which is typically done in active learning.

%\textbf{TODO: add references to other works that use weak labels for training?}

%=======================================================================================================================
% Active learning for sound event detection
%=======================================================================================================================
%\section{Active learning for sound event detection}
\section{Sound event annotation using active learning}
\label{sec:active_learning}

We consider SED tasks where the goal is to predict the presence of a given target event class. The results can also be generalized into the multi-class setting. Given a restricted annotation budget and no initial labels we aim to derive strong labels using active learning to train a SED system. To this end, we propose the following machine guided annotation process.

Let $\mathcal{D}^{(k)}_{L}$ denote the set of labeled audio recordings and $\mathcal{D}^{(k)}_{U}$ the set of unlabeled audio recordings at active learning iteration $k$. Further, let $\mathcal{A}^{(k)} = \{(s^{(j)}_i, e^{(j)}_i, c^{(j)}_i)\}{_{i=1}^{B}}{_{j=1}^{k}}$ denote the annotations of segments, where $s$ denotes the onset, $e$ the offset, and $c\in\{0, 1\}$, the weak label for each segment $i$ of the $B$ annotated segments in audio recording $j$.

We start without any labels, $\mathcal{A}^{(0)} = \mathcal{D}^{(0)}_{L} = \O$, and all audio recordings are unlabeled, $\mathcal{D}^{(0)}_U = \{\mathbf{x}_j\}_{j=1}^{N}$, where $\mathbf{x}_j \in \mathbb{R}^{T}$ denotes an audio recording of length $T$, and $N$ denotes the total number of audio recordings. We then loop for each $k \in \{1, \dots, N\}$ and:
\begin{enumerate}
    \item choose a random unlabeled audio recording $\mathbf{x}$ from $\mathcal{D}^{(k-1)}_U$,
    \item derive a set of $B$ audio query segments $Q=\{q_i\}_{i=0}^{B-1}$ using a query strategy where $q_i = (s_i, e_i)$ consists of the start $s_i$ and end $e_i$ timings for query $i$,
    \item send the queries to the annotator (returning a weak label for each query) and add the annotations to the set of segment labels $\mathcal{A}^{(k)} = \mathcal{A}^{(k-1)} \cup \{(s_i, e_i, c_i)\}_{i=1}^{B}$,
    \item \textit{In case of A-CPD}: use the annotations $\{(s_i, e_i, c_i)\}_{i=1}^{B}$ to update the query strategy, and
    \item update the labeled recording set $\mathcal{D}^{(k)}_L$ by adding $\mathbf{x}$ and the unlabeled recording set $\mathcal{D}^{(k)}_U$ by removing $\mathbf{x}$.
\end{enumerate}
For brevity we have omitted the dependence on $k$ for $\mathbf{x}_{r_k}$ and $(s^{(r_k)}_i, e^{(r_k)}_i, c^{(r_k)}_i)$ in the description of the annotation loop, where $r_k\in\{1, \dots, N\}$ would denote the randomly sampled audio recording for iteration $k$. After the annotation loop all $N$ audio recordings have been annotated exactly once with the query method used in step (2), resulting in a set of annotations $\mathcal{A}^{(N)} = \{(s^{(j)}_i, e^{(j)}_i, c^{(j)}_i)\}{_{i=1}^{B}}{_{j=1}^{N}}$. 

Note that $B$ is not the number of sound events in the recording, but the number of query segments allowed when annotating the recording. The smallest number of query segments to derive the ground truth strong labels does, however, depend on the number of sound events $M$ in the recording as $2M + 1$ (see Section~\ref{sec:oracle_method}). A-CPD is developed to provide strong labels using as few as $B = 2M+1$ queries.

The total annotation budget used will scale with both $N$ and $B$. Typically we would aim to reduce $N$ by actively sampling the data points to annotate, but we instead aim to reduce $B$. Think of $B$ as a part of the annotation cost of an audio recording, which can be reduced with maintained label strength by guiding the annotator during the annotation process.

%=======================================================================================================================
% Query strategies
%=======================================================================================================================
\section{Query strategies}
\label{sec:query_strategies}

%In this section we describe each of the studied query strategies: A-CPD, F-CPD, FIX, and ORC. 

In this section we describe the studied query strategies. %work we study four different query strategies: the proposed query strategy based on adaptive change point detection (A-CPD), a baseline query strategy based on fixed change point detection (F-CPD)~\cite{Shuyang2020} and one based on fixed query lengths (FIX), and a reference oracle query strategy (ORC).

\begin{figure}
    \centering
    \includegraphics[width=\columnwidth]{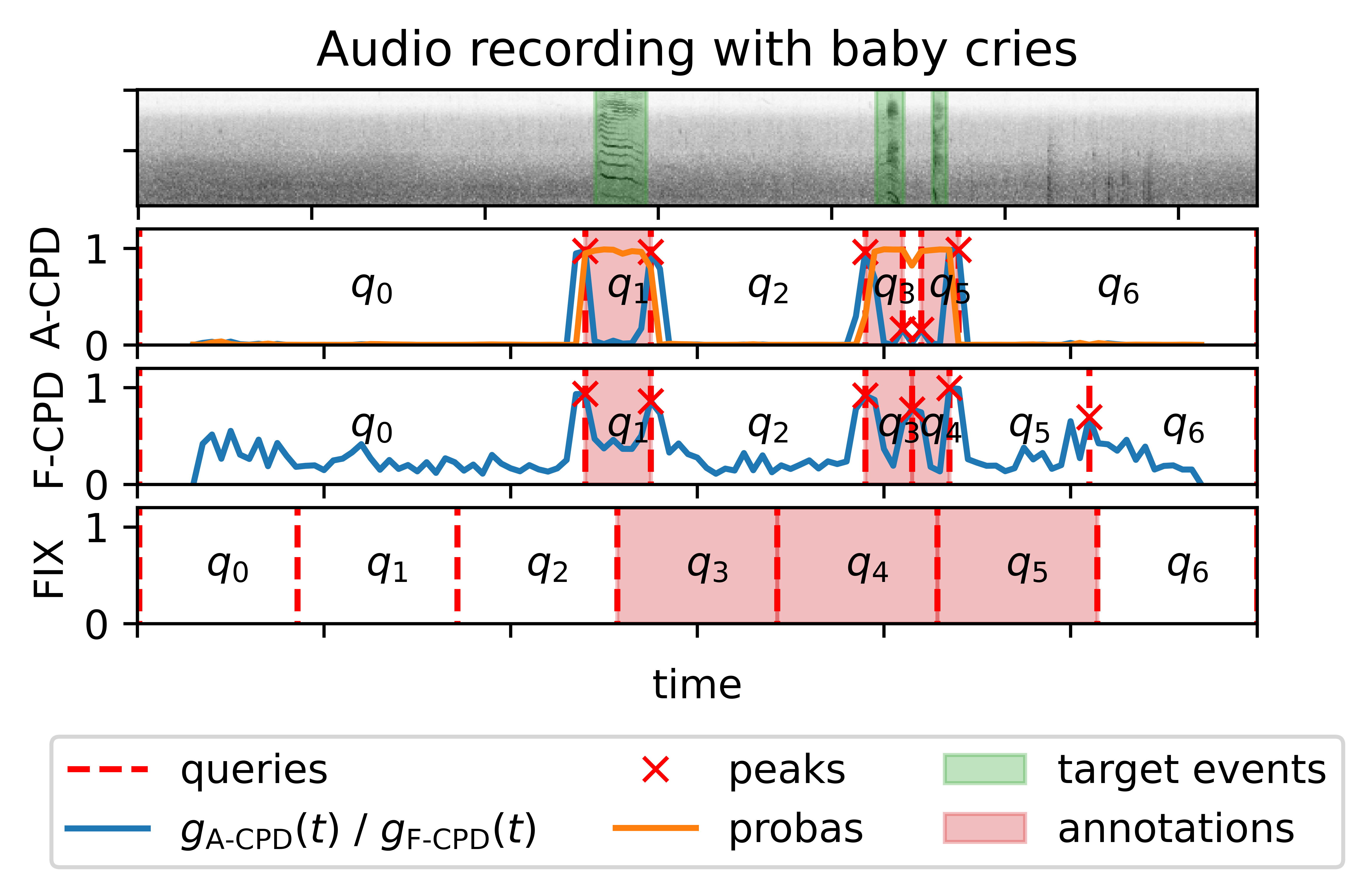}
    \caption{Qualitative example of how the different query strategies A-CPD, F-CPD and FIX segment a spectrogram of an audio recording with three target events shown in shaded green (top panel) into $B=7$ queries. A-CPD (second panel) uses change point detection (blue line) on the probability curve from a prediction model (orange line) to detect the $B-1$ most prominent peaks (red crosses) which are used to construct a set of queries $\{q_0, \dots, q_{B-1}\}$ (dashed red lines). Each query $q_i = (s_i, e_i)$ is given a weak label $c_i \in \{0, 1\}$ ($c=1$ shown as shaded red), resulting in the $i$:th annotation $(s_i, e_i, c_i)$. F-CPD (third panel) uses change point detection directly on the cosine distances in embedding space (blue line) and thereafter constructs queries in the same way as A-CPD. FIX (fourth panel) uses fixed length queries.} %Query $q_4$ for A-CPD is omitted for clarity.}
    \label{fig:method_illustration}
\end{figure}

%------------------------------------------------------------------------------------------------------------------------
% A-CPD
%------------------------------------------------------------------------------------------------------------------------
\subsection{The adaptive change point detection strategy (A-CPD)}

To produce a set of queries for a given audio recording $\mathbf{x}$ at annotation round $k$ we perform three key steps:
\begin{enumerate}
    \item update a prediction model using the annotations from round $k-1$ (initialized with pre-training if $k=0$),
    \item predict probabilities indicating the presence of the target class in the recording using the model, and
    \item apply change point detection to the probabilities to derive the queries.
\end{enumerate}
The pre-training of the prediction model can be done in a supervised or unsupervised way. The important property is that the model reacts to changes in the audio recording related to the presence or absence of the target class. However, it is not strictly necessary that the model reacts \textit{only} to those changes.

Let $h_k : \mathbb{R}^{L} \rightarrow [0,1]$ denote a model that predicts the probability of an audio segment of length $L$ belonging to the target event class. In principle, any prediction model can be used. For a given audio recording $\mathbf{x}$ the prediction model $h_k(\cdot)$ is applied to consecutive audio segments to derive a probability curve, shown as the orange curve for A-CPD in Fig.~\ref{fig:method_illustration}. The consecutive audio segments are derived using a moving window of $L$ seconds with hop size $L/4$.

We define the Euclidean distance between two points $t-\alpha$ and $t+\alpha$ on the probability curve as:
\begin{equation}
\label{eq:A-CPD_F-CPD}
    g^{(k)}_{\text{A-CPD}}(t) = ||h_k(t - \alpha) - h_k(t + \alpha)||,
\end{equation}
shown as the blue curve for A-CPD in Fig.~\ref{fig:method_illustration}.
The previous probability is compared with the next probability in Eq.~\ref{eq:A-CPD_F-CPD}, and $\alpha = L/4$ (hop size) is therefore chosen to ensure a $50$\% overlap between the audio segments for these probabilities.

Let $t$ be a local optimum of $g^{(k)}_{\text{A-CPD}}(t)$, and all such local optima are called peaks. We rank peaks based on \textit{prominence}. For any given peak $t$, let $t_l$ and $t_r$ denote the closest local minima of $g_k(\cdot)$ to the left and right of $t$. The prominence of the peak at $t$ is defined as $|g_k(t) - \max(g_k(t_l), g_k(t_r))|$. Let $\mathcal{T}_{\text{A-CPD}} = \{t_1, t_2, \dots, t_{B-1}\}$ be the $B-1$ most prominent peaks of a given audio recording such that $t_1 \leq t_2 \leq \dots \leq t_{B-1}$, shown as red crosses in Fig.~\ref{fig:method_illustration}. The A-CPD query method is then defined as:
\begin{equation}
\label{eq:a_cpd}
    Q^{(k)}_{\text{A-CPD}} = \{(0, t_1), (t_1, t_2), \dots, (t_{B-1}, T)\},
\end{equation}
which are shown as dashed red lines in Fig.~\ref{fig:method_illustration}, where $T$ is the length of the audio recording and $B$ is the number of queries used. Note that $g^{(k)}_{\text{A-CPD}}(t)$ will gradually become more sensitive towards changes between presence and absence of the target class in the recording with additional annotations, and become less sensitive to other unrelated changes. % between other sound events changes occurring in the recording.

%------------------------------------------------------------------------------------------------------------------------
% F-CPD
%------------------------------------------------------------------------------------------------------------------------
\subsection{The fixed change point detection strategy (F-CPD)}

The fixed change point detection (F-CPD) method used as a reference derives the queries by computing the cosine distance between the previous embedding at time $t-\alpha$ and the next embedding at time $t+\alpha$:
\begin{equation}
\label{eq:g_fix}
    g_{\text{F-CPD}}(t) = 1 - \frac{\mathbf{e}_{t - \alpha} \cdot \mathbf{e}_{t + \alpha}}{||\mathbf{e}_{t - \alpha}||||\mathbf{e}_{t + \alpha}||},
\end{equation}
where $\mathbf{e}_t = f_{\theta}(\mathbf{x}_t)$ denotes the embedding of consecutive audio segments $\mathbf{x}_t$ centered at second $t$ using the embedding function $f_{\theta} : \mathbb{R}^L \rightarrow \mathbb{R}^K$. 
%The embedding function embeds the audio into a structurally meaningful embedding space of dimension $K$. 
The cosine distance curve for an audio recording is shown as the blue line for F-CPD in Fig.~\ref{fig:method_illustration}. This method is similar to~\cite{Shuyang2020} except that embeddings are derived for $1.0$ seconds of audio instead of $0.02$. We therefore directly compare the previous and next embeddings instead of a moving average as in~\cite{Shuyang2020}.

The most prominent peaks in the cosine distance curve is then selected, $\mathcal{T}_{\text{FIX}} = \{t_1, t_2, \dots, t_{B-1}\}$, and the set of queries are defined as in Eq.~\ref{eq:a_cpd}, shown as dashed red lines for F-CPD in Fig~\ref{fig:method_illustration}.

%------------------------------------------------------------------------------------------------------------------------
% FIX
%------------------------------------------------------------------------------------------------------------------------
\subsection{The fixed length strategy (FIX)}
In the fixed length query strategy (FIX) audio is split into equal length segments and then labeled. Let $d = T / B$, then the queries are defined as
\begin{equation}
\label{eq:fix}
    Q_{\text{FIX}} = \{ (0d, 1d), (1d, 2d), \dots, ((B-1)d, Bd)\},
\end{equation}
shown as dashed red lines for FIX in Fig~\ref{fig:method_illustration}. This is the setting most previous active learning work for SED consider. 

%------------------------------------------------------------------------------------------------------------------------
% ORC
%------------------------------------------------------------------------------------------------------------------------
\subsection{The oracle strategy (ORC)}
\label{sec:oracle_method}
The oracle query strategy constructs the queries based on the ground truth presence and absence annotations
\begin{equation}
\label{eq:orc}
    Q_{\text{ORC}} = \{(s_0, e_0), (s_1, e_1), \dots, (s_{B_{\text{suff}}-1}, e_{B_{\text{suff}}-1})\},
\end{equation}
where $(s_i, e_i)$ is the onset and offset for segment $i$ where the target event is either present or not. $B_{\text{suff}}$ is the sufficient number of queries to get the true strong labels, which relate to the number of target events $M$ in the given audio recording by $B_{\text{suff}} = 2M + 1$. ORC is undefined for $B < B_{\text{suff}}$. %, where $M$ is the number of events.

\subsection{The role of query strategies in the annotation process}

The query strategies described in this section are then used in step (2) of the annotation loop described in Section~\ref{sec:active_learning}. Note that when the queries are not adapted to the audio recording multiple events can end up being counted as one. In Fig.~\ref{fig:method_illustration} we can see this for F-CPD where $q_3$ and $q_4$ are directly adjacent, meaning that they are not resolved as two separate events, and for FIX where $q_3$, $q_4$ and $q_5$ are all directly adjacent. A-CPD often resolves all three events. Fig~\ref{fig:method_illustration} is a qualitative example of all three methods, and quantitative results to further support this claim are provided later in table~\ref{tab:train_without_noise}.

The FIX length query segments depend on the query timings and target event timings aligning by chance since the query construction is independent of the target events. The A-CPD method aim to create query segments that are aligned with the target events by construction. In addition, the number of queries needed to derive the strong labels scale with the number of target events in the recording for A-CPD, which can be beneficial.

%instead of the audio duration and the event lengths as for FIX. %The number of queries needed to find strong labels using the FIX method scales with the with the total duration of the audio to be annotated $TN$ divided by the average target event duration, which means that finding short target events in long-term audio recordings requires many queries. The number of queries needed to find strong labels using the A-CPD method scales with the number of target events, which is typically a much smaller number when looking for short target events.

%=======================================================================================================================
% Evaluation
%=======================================================================================================================
\section{Evaluation}
\label{sec:evaluation}

%------------------------------------------------------------------------------------------------------------------------
% Datasets
%------------------------------------------------------------------------------------------------------------------------
\subsection{Datasets}
\label{sec:datasets}

We create three SED datasets for evaluation, each with a different target event class: Meerkat, Dog or Baby cry. The Meerkat sounds are from the DCASE 2023 few-shot bioacoustic SED dataset~\cite{Nolasco2022} and the Dog and Baby cry sounds from the NIGENS dataset~\cite{Trowitzsch2020}. The sounds used for absence of an event are from the $15$ background types in the TUT Rare sound events dataset~\cite{Diment2018}.

The audio recordings in each dataset are created by randomly selecting $M=3$ sound events from that event class and mixing them together with a randomly selected background recording of length $T=30$ seconds. In this way we know that exactly $B_{\text{suff}}=2M+1=7$ queries are \textit{sufficient} and \textit{necessary} to derive the ground truth strong labels using a weak label annotator. The mixing is done using Scaper~\cite{Salamon2017} at an SNR of $0$ dB. In total we generate $N=300$ audio recordings using this procedure for each event class as training data and equally many as test data.

The source files used in the mixing uses the supplied splits in~\cite{Trowitzsch2020} and~\cite{Diment2018}, except for the Meerkat sounds where non exist and the split is done on a recording level.

%------------------------------------------------------------------------------------------------------------------------
% Evaluation metrics
%------------------------------------------------------------------------------------------------------------------------
\subsection{Evaluation metrics}
\label{sec:metrics}
We evaluate the methods by annotating the mixed training datasets using the query strategies described in Section~\ref{sec:active_learning} and the annotation loop described in Section~\ref{sec:query_strategies}. The quality of the annotations are then measured in two ways: (i) how strong the annotations are compared to the ground truth, and (ii) the test time performance of two evaluation models trained using the different annotations. 

The evaluation metrics used in case (i) and (ii) are event-based $F_1$-score ($F_{1e}$) and segment-based $F_1$-score ($F_{1s}$)~\cite{Mesaros2016}. The segment size for $F_{1s}$ is set to $0.05$ seconds, and the collar for $F_{1e}$ is set to $0.5$ seconds. In case (i) the $F_{1s}$ measures how much of the audio that has been correctly labeled and in case (ii) $F_{1s}$ measures how much of the audio that has been correctly predicted by the evaluation model. The $F_{1e}$ score is only used to measure how close the annotations are to the ground truth labels in the training data.

%Note that it is not possible to derive the true labels using $B < B_{\text{suff}}$ queries, making ORC undefined.

%------------------------------------------------------------------------------------------------------------------------
% Annotator model and cost
%------------------------------------------------------------------------------------------------------------------------
\paragraph{Annotator model}
\label{sec:annotator_model}

Let $\mathcal{A}^{(j)}_{gt} = \{(s_i, e_i, c=1)\}{_{i=1}^{3}}$ denote the set of ground truth target event labels for audio recording $j$, where $s_i$ is the onset, $e_i$ the offset and $c=1$ indicate the presence of the target event. %We consider audio recordings that have exactly three target events, i.e., $M_j = 3$ for all audio recordings $j$.

We use $\mathcal{A}^{(j)}_{gt}$ to simulate an annotator for recording $j$. For a given query segment we check the overlap ratio with the ground truth target event labels. Formally, if there exists an annotation $(s_i, e_i, c_i=1)$ s.t.
\begin{equation}
    \frac{(s_i, e_i) \cap (s_q, e_q)}{|s_i-e_i|} \geq \gamma,
\end{equation}
holds for the given query segment $q = (s_q, e_q)$, then the annotator returns $c_i = 1$ for query $q$, and $c_i = 0$ otherwise. Annotation noise is added by flipping the returned label with probability $\beta$. In this work $\gamma=0.5$, and $\beta\in\{0.0, 0.2\}$.

%------------------------------------------------------------------------------------------------------------------------
% Implementation details
%------------------------------------------------------------------------------------------------------------------------
\subsection{Implementation details and experiment setup}
\label{sec:implementation}

\paragraph{Prediction model}
The prediction model $h_k(\cdot)$ is modeled using a prototypical neural network (ProtoNet)~\cite{Snell2017}. The prototypes are easily updated at each annotation round $k$ using a running average between each previous prototype and the newly labeled audio embeddings. We model the embedding function $f_{\theta}(\cdot)$ using BirdNET~\cite{Kahl2021}, a convolutional neural network pre-trained on large amounts of bird sounds.

\paragraph{Evaluation models} We use two models to evaluate the test time performance of models trained on the annotations obtained using each query strategy: a two layer multilayer perceptron (MLP) and a ProtoNet. 
%The MLP consists of a batch normalization followed by two linear layers with ReLU activations of size $256$ and $2$. 
The MLP is trained using the Adam optimizer and cross-entropy loss.
%with a learning rate of $3\times 10^{-4}$.
Each query strategy is run $10$ times and the evaluation models are trained on the embeddings using the resulting labeled datasets. ProtoNet is used in two ways: as a prediction model in the proposed A-CPD method, and as an evaluation model. % to measure test performance.

%------------------------------------------------------------------------------------------------------------------------
% Results
%------------------------------------------------------------------------------------------------------------------------
\subsection{Results}
\label{sec:results}

In Table~\ref{tab:train_without_noise} we show the average $F_{1s}$-score and $F_{1e}$-score for the training data annotations over $10$ runs for each dataset and with the sufficient nu $B=B_{\text{suff}}=7$. The A-CPD method outperforms the other methods for all studied target event classes. The standard deviation is in all cases less than $0.03$ (omitted from table for brevity), and the baseline query strategies are deterministic when $\beta = 0$.% i.e., there is no deviation at all.

\begin{table}[]
    \caption{Average $F_{1s}$-score and $F_{1e}$-score for the training annotations for each annotation process and target event class with $\beta = 0$}
    \centering
    \begin{tabular}{c |cc |cc|cc}
Strategy & \multicolumn{2}{|c|}{Meerkat} & \multicolumn{2}{|c|}{Dog} & \multicolumn{2}{|c}{Baby} \\
         & $F_{1s}$ & $F_{1e}$               & $F_{1s}$ & $F_{1e}$           & $F_{1s}$ & $F_{1e}$ \\
\hline
ORC & $1.00$ & $1.00$ & $1.00$ & $1.00$ & $1.00$ & $1.00$ \\
\hline
A-CPD & $\mathbf{0.31}$ & $\mathbf{0.57}$ & $\mathbf{0.29}$ & $\mathbf{0.45}$ & $\mathbf{0.62}$ & $\mathbf{0.60}$ \\
F-CPD & $0.16$ & $0.44$ & $0.21$ & $0.30$ & $0.48$ & $0.45$ \\
FIX & $0.11$ & $0.00$ & $0.19$ & $0.00$ & $0.41$ & $0.01$ \\
    \end{tabular}

    \label{tab:train_without_noise}
\end{table}

\begin{figure}
    \centering
    \includegraphics[width=1.0\columnwidth]{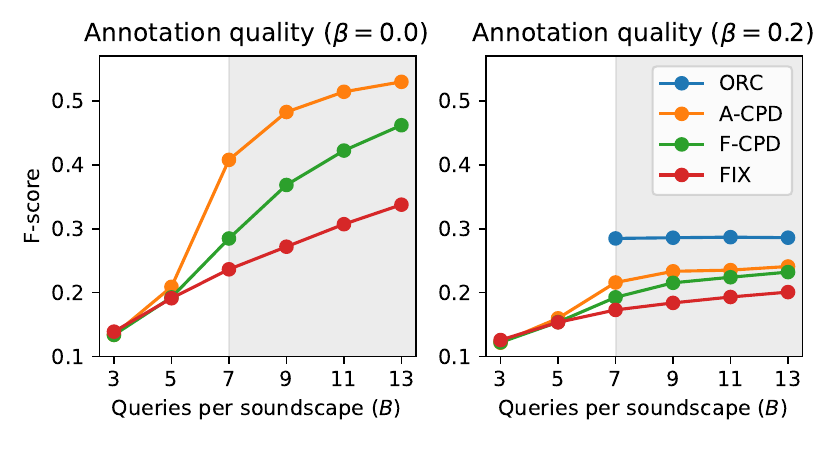}
    \caption{The average $F_{1s}$-score over the three classes for each of the studied annotation processes plotted against the number of queries per audio recording, $B$. The results are shown for an annotator without noise (left) and with $\beta=0.2$ (right). Note that ORC is $1.0$ when $\beta=0$ and is therefore not shown in the left figure. Shaded region where $B \geq B_{\text{suff}}$.}
    \label{fig:train}
\end{figure}

In Fig.~\ref{fig:train} we show the average $F_{1s}$-score over all runs and event classes for the annotations derived from each query strategy. The proposed A-CPD method has a strictly higher $F_{1s}$-score than the FIX and F-CPD baselines for all budgets and noise settings. We also see that there is still a significant gap to the ORC strategy. The noisy annotator ($\beta=0.2$) drastically reduce the label quality for all studied strategies, especially ORC dropping from an $F_{1s}$-score of $1.0$ (omitted from figure) to $\approx 0.28$ (large drop due to class-imbalance). %We have shaded the region where $B > B_{\text{suff}}$ in all plots.

\begin{figure}
    \centering
    \includegraphics[width=1.0\columnwidth]{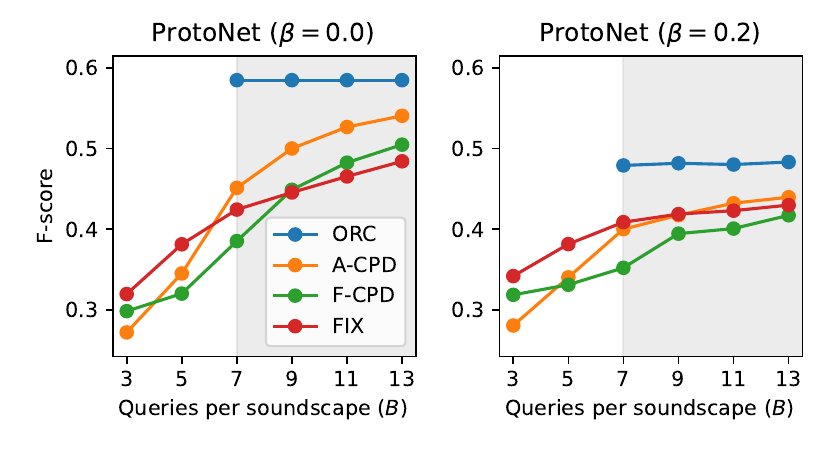}
    \includegraphics[width=1.0\columnwidth]{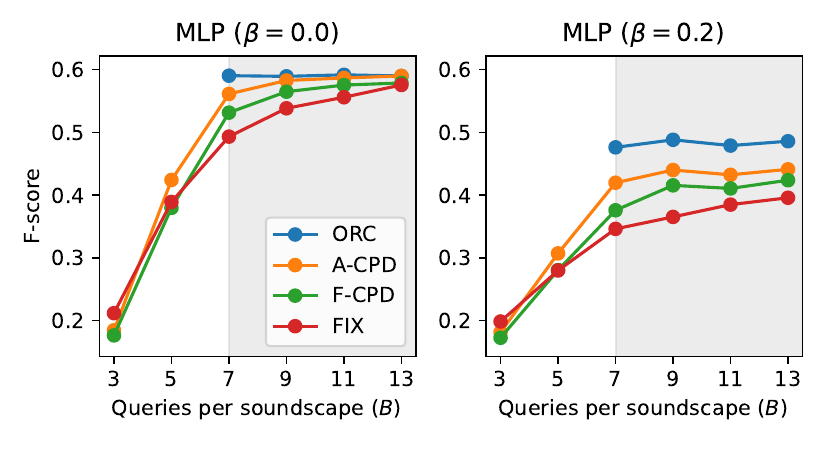}
    \caption{The average test time $F_{1s}$-score over the studied sound classes for a ProtoNet (top) and the MLP (bottom) trained with the annotations from each respective annotation process and setting. Shaded region where $B \geq B_{\text{suff}}$.}
    \label{fig:test_performance}
\end{figure}

In Fig.~\ref{fig:test_performance} we show the average test time $F_{1s}$-score of a ProtoNet (top) and a MLP (bottom) trained using the annotations from each of the studied annotation strategies and settings. The A-CPD method outperforms the other methods when $B\geq 7$. For the ProtoNet the FIX method outperform A-CPD when $B < 7$ and for the MLP the results are similar. 

%Fig.~\ref{fig:test_mlp} shows the average segment-based test F-sore of the MLP. The A-CPD method outperforms all other methods in all budget settings and noise settings except when $B=3$.

%\begin{figure}
%    \centering
%    \includegraphics[width=1.0\columnwidth]{figures/mlp_test.pdf}
%    \caption{The average segment-based test F-score over the studied sound classes for a MLP trained with the annotations %from each respective annotation process and setting.}
%    \label{fig:test_mlp}
%end{figure}

%%%%%%%%%%%%%%%%%%%%%%%%%%%%%%%%%%%%%%%%%%%%%%%%%%%%%%%%%%%%%%%%%%%%%%%%%%%%%%%%%%%%%%%%%%
% prototypical test performance
%%%%%%%%%%%%%%%%%%%%%%%%%%%%%%%%%%%%%%%%%%%%%%%%%%%%%%%%%%%%%%%%%%%%%%%%%%%%%%%%%%%%%%%%%%

Table~\ref{tab:proto_test_without_noise} and~\ref{tab:mlp_test_without_noise} show the average $F_{1s}$-score and standard deviation for the three different event classes for all studied query strategies. The average is over $10$ runs, and the number of queries is set to $B=7$. Table~\ref{tab:proto_test_without_noise} shows the $F_{1s}$-score for the ProtoNet evaluation model. A-CPD achieves a higher $F_{1s}$-score for the meerkat and baby datasets. On average A-CPD outperforms the other methods as seen in Fig.~\ref{fig:test_performance}. Table~\ref{tab:mlp_test_without_noise} shows the $F_{1s}$-score for the MLP evaluation model. A-CPD achieves a higher $F_{1s}$-score for all studied datasets.

\begin{table}[]
    \centering
    \caption{Average test time $F_{1s}$-score for ProtoNet with $\beta = 0$.}
    \begin{tabular}{c | c | c | c |}

Strategy & Meerkat & Dog & Baby \\
\hline
ORC & $0.46$ & $0.48$ & $0.81$ \\
\hline
A-CPD & $\mathbf{0.44} \pm 0.00$ & $0.20 \pm 0.01$ & $\mathbf{0.71} \pm 0.02$ \\
F-CPD & $0.31$          & $0.19$          & $0.66$ \\
FIX   & $0.34$          & $\mathbf{0.25}$          & $0.68$ \\
    \end{tabular}
    
    \label{tab:proto_test_without_noise}
\end{table}

%%%%%%%%%%%%%%%%%%%%%%%%%%%%%%%%%%%%%%%%%%%%%%%%%%%%%%%%%%%%%%%%%%%%%%%%%%%%%%%%%%%%%%%%%%
% mlp test performance
%%%%%%%%%%%%%%%%%%%%%%%%%%%%%%%%%%%%%%%%%%%%%%%%%%%%%%%%%%%%%%%%%%%%%%%%%%%%%%%%%%%%%%%%%%

%Note that the F-score of the ProtoNet using the ORC labeling process (Table~\ref{tab:proto_test_without_noise}) is smaller than the F-score of the MLP using the ORC labels (Table~\ref{tab:mlp_test_without_noise}).

 %Note that on average A-CPD outperforms the other methods as seen in Fig.~\ref{fig:test_performance}.

\begin{table}[]
    \centering
    \caption{Average test time $F_{1s}$-score for MLP with $\beta = 0$.}
    \begin{tabular}{c | c | c | c |}

Strategy & Meerkat & Dog & Baby \\
\hline
ORC   & $0.43 \pm 0.00$ & $0.51 \pm 0.01$ & $0.83 \pm 0.00$ \\
\hline
A-CPD & $\mathbf{0.44} \pm 0.00$ & $\mathbf{0.43} \pm 0.02$ & $\mathbf{0.81} \pm 0.01$ \\
F-CPD & $0.38 \pm 0.01$ & $0.42 \pm 0.02$ & $0.79 \pm 0.01$ \\
FIX   & $0.33 \pm 0.02$ & $0.40 \pm 0.02$ & $0.75 \pm 0.02$ \\
    \end{tabular}
    
    \label{tab:mlp_test_without_noise}
\end{table}

%=======================================================================================================================
% Discussion
%=======================================================================================================================
\subsection{Discussion}
\label{sec:discussion}

The results in all tables are for the sufficient budget $B = B_{\text{suff}} = 2M + 1$. In practice we do not know $B_{\text{suff}}$. However, the A-CPD method is applicable also for an arbitrary number of sound events in the recording when $B$ is chosen sufficiently large. This choice need to be made for all the studied methods. We show the benefit of A-CPD for differently chosen $B$ in Fig.~\ref{fig:train}. Estimating $B_{\text{suff}}$ based on the audio recording could further reduce the number of queries used and is left as future work.

We chose $\gamma = 0.5$ in the annotator model since the annotator should be able to detect a target event if more than $50$\% of the event occurs within the query segment. This choice is however non-trivial, and depends on the expertise of the annotator and target class among others. We observe similar results on average as those presented in the paper for $\gamma \in \{0.05, 0.25, 0.5, 0.75, 0.95\}$ (not shown).

We use BirdNET~\cite{Kahl2021} to model the embedding function since we study bioacoustic target classes. However, an embedding function such as PANNs~\cite{Kong2020a} may also be used if the target classes are more general.

%An important choice for both A-CPD and F-CPD is the window size $L$ used for the embeddings. This directly affect the temporal resolution of the change point detection. Looking at how to best choose $L$ is relevant future work. %Future work will look at how to choose this parameter in a principled way based on either prior knowledge of the event length distribution or learn it as a part of the annotation loop.

%\textbf{TODO: other embedding functions like PANNs and YamNET}

%=======================================================================================================================
% Conclusions
%=======================================================================================================================
\section{Conclusions}
\label{sec:conclusions}
We have presented a query strategy based on adaptive change point detection (A-CPD) which derive strong labels of high quality from a weak label annotator in an active learning setting. We show that A-CPD gives strictly stronger labels than all other studied baseline query strategies for all studied budget constraints and annotator noise settings. We also show that models trained using annotations from A-CPD tend to outperform models trained with the weaker labels from the baselines at test time. We note that the gap to the oracle method is still large, leaving room for improvements in future work.

\bibliographystyle{IEEEtran}
\bibliography{refs}

\end{document}